\documentclass[a4paper,11pt]{article}
\usepackage{jinstpub} 
\usepackage{graphicx}
\usepackage{sidecap}
\usepackage{wrapfig}
\usepackage{float}
\usepackage{supertabular}
\usepackage{array}
\usepackage{threeparttable}
\usepackage{booktabs}
\usepackage[normalem]{ulem}
\usepackage{setspace}
\usepackage{url}
\usepackage{amsmath}
\usepackage{amsfonts}
\usepackage{indentfirst}
\usepackage{subcaption}
\usepackage{caption}
\usepackage{pdflscape}
\usepackage{lipsum}
\usepackage{blindtext}
\usepackage[nottoc]{tocbibind}
\usepackage{pgfplots}
\usepackage{pgfplotstable}
\usepackage{multirow}
\usepackage{tikz}
\usepackage{xcolor}
\usepackage{etoolbox}
\usepackage{flafter}
\usepackage{placeins}
\usepackage{booktabs} 
\usepackage{capt-of} 
\usepackage{graphicx} 
\usepackage{cleveref}
\usepackage{algorithm}
\usepackage{algpseudocode}

\title{\centering \huge MEDIC: a network for monitoring data quality in collider experiments}

\author{Juvenal Bassa,}
\author[1]{Arghya Chattopadhyay\note{Corresponding author.},}
\author{Sudhir Malik}
\author{and Mario Escabi Rivera}

\smallskip\vfill
\affiliation{Department of Physics, University of Puerto Rico Mayaguez, PR 00681 USA}
\smallskip\vfill
\emailAdd{juvenal.bassa@upr.edu}
\emailAdd{arghya.chattopadhyay@upr.edu}
\emailAdd{sudhir.malik@upr.edu}
\emailAdd{mario.escabi@upr.edu}

\smallskip \vfill
\abstract{Data Quality Monitoring (DQM) is a crucial component of particle physics experiments and ensures that the recorded data is of the highest quality, and suitable for subsequent physics analysis. Due to the extreme environmental conditions, unprecedented data volumes, and the sheer scale and complexity of the detectors, DQM orchestration has become a very challenging task. Therefore, the use of Machine Learning (ML) to automate anomaly detection, improve efficiency, and reduce human error in the process of collecting high-quality data  is unavoidable. Since DQM relies on real experimental data, it is inherently tied to the specific detector substructure and technology in operation. In this work, a simulation-driven approach to DQM is proposed, enabling the study and development of data-quality methodologies in a controlled environment. Using a modified version of Delphes -- a fast, multi-purpose detector simulation -- the preliminary realization of a framework is demonstrated which leverages ML to identify detector anomalies as well as localize the malfunctioning components responsible. We introduce \textbf{MEDIC} (Monitoring for Event Data Integrity and Consistency), a neural network designed to learn detector behavior and perform DQM tasks to look for potential faults. Although the present implementation adopts a simplified setup for computational ease, where large detector regions are deliberately deactivated to mimic faults, this work represents an initial step toward a comprehensive ML-based DQM framework. The encouraging results underline the potential of simulation-driven studies as a foundation for developing more advanced, data-driven DQM systems for future particle detectors.}

\begin{document}
	
	\maketitle
	\flushbottom 
	
	\section{Introduction}\label{sec:intro}
	Validation of any scientific theory in physics fundamentally depends on the reliability of experimental data. The more complex an experiment is, the harder it becomes to ensure the integrity of that data. Particle detectors represent the pinnacle of engineering and experimental physics, but they also carry the immense challenge of maintaining error free data acquisition. At the scale of experiments such as the Large Hadron Collider (LHC) at CERN (the European Organization for Nuclear Research), where vast amounts of data are collected continuously, a robust and efficient data quality monitoring (DQM) system is indispensable. Each of the experimental apparatus at the LHC \cite{CMS:2008xjf, ATLAS:2008xda, LHCb:2008vvz, ALICE:2008ngc} requires own specific DQM analysis due to their unique and complex multi-component structures \cite{DQM_CMS, DQM_ATLAS_on, DQM_ATLAS_off, DQM_LHCb, DQM_ALICE}. In the following, we will use the CMS detector as a prototypical example whenever required, however all statements made will be general enough to apply to any other experiment.\\

	Traditionally, DQM systems are validated by human shifters reviewing the data collected from various detector components. The primary target of a DQM system is to represent the collected data in a format (most often as sets of histograms) such that it becomes easy for the shifter to compare it with the expected or reference output. These reference outputs are constructed manually by human experts from previously validated datasets. To enforce stringent quality criterion and optimize time and effort, DQM operations are typically divided into two main domains:

	\begin{itemize}
		\item \textit{Online monitoring:} This is the first step in the DQM chain, where shifters review the live feedback from the detector while the data is being taken. This stage enables rapid identification and resolution of issues. Data rejected at this step will not be processed for further analysis \cite{CMS:2018qpa, ATLAS:2019fst}.
		
		\item \textit{Offline monitoring:} Also referred to as the \emph{data certification}  step, this phase involves a thorough review and certification of the collected data before it is used for physics analysis \cite{Kalsi:20208L}.
	\end{itemize}
	A crucial distinction between these two steps lies in the volume of data being analyzed. Since online monitoring needs to be performed almost at real-time scale, it uses only a smaller fraction of the dataset, whereas a robust analysis over a larger dataset is performed during the offline phase.\\

	Given the complexity of the particle detectors, resolving detector-level failures is an inevitable part of any collider operation. Because of the built-in redundancies in the detectors, the relevant physics can often still be extracted even if some local components of the detectors under perform. Nevertheless, as part of good scientific practice, all imperfections in the detector during a run must be recorded. Therefore, an effective DQM system should not only be sensitive to failures but also specific enough to identify the precise source of the issue. The ever increasing complexity of the detectors and the sheer volume of monitoring data necessitate new methods to be implemented in high-energy physics (HEP) monitoring, as the traditional approaches are reaching their limits. In recent years, several machine learning (ML) models have already been introduced to assist the human shifters \cite{pol:hal-03159873, CMSECAL:2023fvz, Brinkerhoff:2025rob, Gavrikov:2025tzc}, gradually steering the field of DQM toward increasing automation. \\

	The main objective of this paper is to introduce an end-to-end simulation-driven DQM framework. In a fully developed DQM analysis, readout information from each detector component must ideally be incorporated to capture the full spectrum of possible malfunctions. As an initial step towards that realization, we have modified\footnote{At the time of writing this paper this feature has already been included as a part of the standard Delphes installation, but for a detailed outline of the required modification, the reader can follow the Delphes fork at \cite{arghya_delphes}.} the \textbf{Delphes} fast simulation framework \cite{deFavereau:2013fsa}, to simulate various categories of detector malfunctions. Being a fast simulation, Delphes generates data at the level of reconstructed particle kinematics rather than raw detector signals, the details of which are discussed in the following section. Traditional DQM approaches primarily rely on statistical consistency tests such as the $\chi^2$ or the Kolmogorov–Smirnov tests to compare histograms against reference datasets \cite{Piparo:2012gm}. In contrast, ML-driven DQM methods involve supervised classifiers, often convolutional neural networks (CNNs) or autoencoder-based architectures \cite{pol:hal-03159873, CMSECAL:2023fvz}, or a hybrid approach of combining both statistical techniques and unsupervised ML \cite{Brinkerhoff:2025rob} or transformer based attention models \cite{Gavrikov:2025tzc} or transfer learning methods \cite{Asres_2025} to identify anomalies in detector read-outs. \\

	Contrary to these histogram-based approaches, we introduce \textbf{MEDIC} (\textbf{M}onitoring for \textbf{E}vent \textbf{D}ata \textbf{I}ntegrity and \textbf{C}onsistency), a DQM framework that directly utilizes the particle-level kinematic information from simulated detector outputs. MEDIC is trained to classify events into distinct categories corresponding to specific detector malfunctions. Since MEDIC operates directly on raw detector-level kinematic inputs without any intermediate histograms or manual feature extraction, the entire classification process, from detector output to anomaly identification, is fully end-to-end. This event-level perspective allows the network to identify subtle inconsistencies that may not otherwise be visible in aggregated histogram data, thereby improving latency and sensitivity to detector anomalies.\\

	One of the major advantages of employing a simulation-driven approach is that the reference run, representing the detector’s ideal, glitch-free condition, can be generated trivially through the same simulation chain. This capability enables a self-consistent and reproducible DQM framework that does not depend on pre-certified data from actual detector runs. Without such a ground-up simulation strategy, one could alternatively inject random errors into certified data and train an ML model to detect anomalies. While these might teach a model to recognize that an anomaly exists, identifying its physical origin would still require intervention from a human expert. The present approach eliminates this need for manual intervention by simulating realistic detector glitches directly, rather than introducing random perturbations. Moreover, this approach is particularly valuable in the context of the High-Luminosity LHC (HL-LHC) upgrade \cite{Apollinari:2015wtw}, where substantial detector modifications and new failure modes are expected. Instead of waiting for certified data to retrain DQM models, simulation-based reference datasets can be updated immediately to reflect hardware or configuration changes. This flexibility not only reduces latency in the data certification process (both online and offline) but also enables continuous DQM development, allowing the monitoring algorithms to evolve in parallel with detector upgrades. The complete MEDIC software pipeline, including simulation configurations and anomaly-generation parameter cards for Delphes simulations are available publicly in the GitHub repository \cite{medic}. This open framework allows reproducibility and easy adaptation to other LHC experiments or future colliders.\\

	The remainder of this paper is organized as follows: In \cref{sec:det_sim}, the details of detector simulations and the procedure used to generate different categories of detector anomalies are discussed. \Cref{sec:train_test} outlines the architecture of the MEDIC network, the training strategy, and the performance metrics used to evaluate its accuracy. \Cref{sec:results} presents the main results of training and testing of the MEDIC network, followed by a discussion of the implications and future directions in \cref{sec:discussion}. Finally, Appendix \ref{app:datas_create} provides technical details of the dataset preparation.

	\section{Detector simulation with irregularities}\label{sec:det_sim}
	Ideally, a full fledged DQM analysis requires both reconstructed particle kinematics and highly granular information from the underlying detector subsystems. Although such electronic level detail can also be generated using a full simulation through Geant4 based detector simulators \cite{GEANT4:2002zbu}, we kept that computationally intensive approach for future purposes. As a first step toward this broader goal, we adopt Delphes \cite{deFavereau:2013fsa}, a fast-simulation framework that provides realistic but higher-level detector responses. While Delphes does not model low-level electronics or detailed shower development, it allows the user to define the geometry, segmentation, and resolution of each detector subsystem such as the tracker, electromagnetic calorimeter (ECAL), and hadronic calorimeter (HCAL) to emulate the behavior of a realistic experimental setup. The simulation chain takes care of effects of the magnetic field and each of the sub-detector resolutions as well as detector smearing to produce a ROOT library for visualization. In this study, the response of the Compact Muon Solenoid (CMS) detector was modeled using the \texttt{SimpleCalorimeter} module implemented in Delphes. We further extended this module by introducing the option to define \texttt{InsensitiveEtaPhiBins} \cite{arghya_delphes}, enabling the simulation of localized detector malfunctions by deactivating any specific regions of choice in the calorimeter. This feature allows controlled emulation of potential readout failures, which in real detectors can arise from any unplanned electrical or hardware issues.\\

	For positional details of the detector components, Delphes uses the traditional coordinate system of collider physics. With three-dimensional Cartesian coordinates, the $z$-axis is defined as the direction of the beam axis and the transverse momenta of the particles along the $(x,y)$ plane as $p_T$. The kinematic information for each particle detected can then be stored as $(p_T,\eta, \phi)$ coordinate along with other details, where $\phi$ is calculated as the angle of the particle trajectory with respect to the $x$ axis and $\eta=-\log(\tan(\theta/2))$ is the pseudo-rapidity with $\theta$ being the polar angle. Delphes collects calorimetric information from both ECAL and HCAL as tower-based observables, which are used as input features in this analysis. In addition to particle kinematics, Delphes also provides global event-level quantities such as the missing transverse energy ($E_T^{\text{miss}}$), which represents the imbalance of transverse momentum in the event and is a key observable for processes involving invisible particles such as neutrinos or potential dark matter candidates.\\
	\begin{figure}[ht]
		\centering
		\begin{subfigure}[b]{0.45\textwidth}
			\centering
			\includegraphics[width=\textwidth]{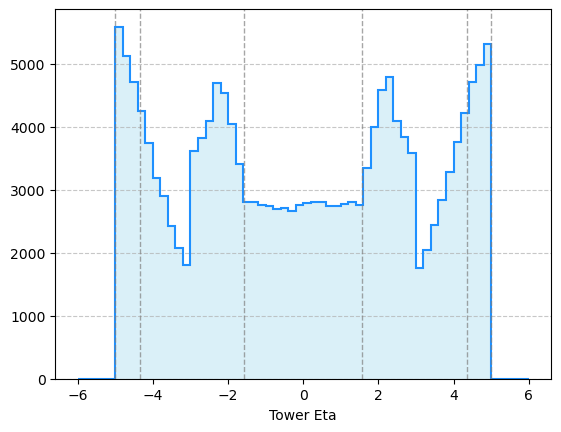}
			\caption{Reference run with all systems intact.}
			\label{subfig:example_zee_norm}
		\end{subfigure}
		\hfill
		\begin{subfigure}[b]{0.45\textwidth}
			\centering
			\includegraphics[width=\textwidth]{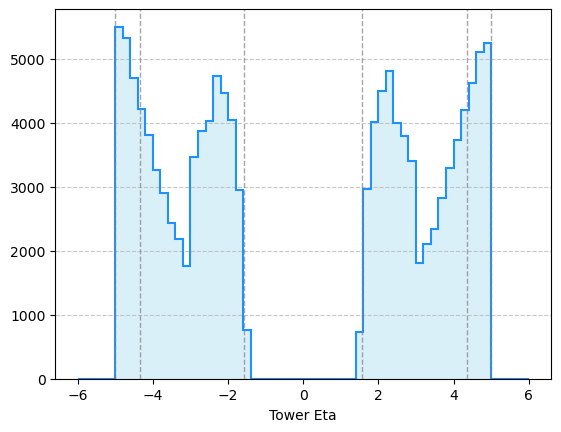}
			\caption{Both ECAL and HCAL barrels deactivated.}
			\label{subfig:example_zee_norm_barell}
		\end{subfigure}
		
		\caption{Pedagogic illustration of tower eta distribution for different scenarios.}
		\label{fig:example_zee}
	\end{figure}

	For the present simulation, the ECAL was configured with a granularity of $\Delta\eta \times \Delta\phi = 0.02 \times 0.02$ in both the barrel ($|\eta| < 1.5$) and endcap regions ($1.5 < |\eta| < 3.0$), while the forward region ($3.0 < |\eta| < 5.0$) followed the CMS calorimeter segmentation setup \cite{Penzo:2009zz}. The corresponding energy resolution and performance parameters were chosen to be consistent with the official CMS detector parameterizations\footnote{The complete Delphes configuration cards, including the ECAL and HCAL resolution parameters, are available in the associated GitHub repository \cite{medic}.} \cite{CMS:2013lxn, CMS:2015xaf}. The HCAL was modeled with non-uniform segmentation corresponding to three tower configurations: $5^\circ$ towers covering the barrel region ($|\eta| \lesssim 1.6$), $10^\circ$ towers for the forward regions ($1.6 \lesssim |\eta| \lesssim 4.3$), and $20^\circ$ towers for the very forward calorimeter ($4.3 < |\eta| \leq 5.0$). The minimum energy thresholds for HCAL towers were set to $1.0$ GeV with \texttt{SmearTowerCenter} set to \texttt{true} to simulate finite spatial resolution effects. By defining  \texttt{InsensitiveEtaPhibBins}, one can then switch off any part of choice or even the entirety of the detector components. To illustrate the procedure,  $Z\rightarrow ee$ events\footnote{with the dataset from \url{http://cp3.irmp.ucl.ac.be/downloads/z_ee.hep.gz}.} 
	are simulated purely for pedagogic completeness as an example with both the ECAL and HCAL for the barrel regions of the CMS detector being \emph{turned off}. This illustrative configuration is not intended to represent the experimental setup used in the present work, but rather to provide a clear visual demonstration of the effect of insensitive detector regions. For a DQM analysis, one can then use \cref{subfig:example_zee_norm} as the reference run and \cref{subfig:example_zee_norm_barell} as the anomalous run where no particles are located near the barrel region of the CMS detector.\\
	\begin{figure}[ht]
		\centering
		\begin{subfigure}{0.45\textwidth}
			\centering
			\includegraphics[width=\linewidth]{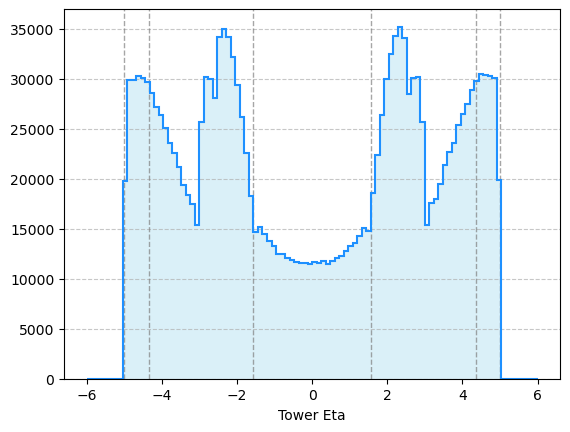}
			\caption{Distribution for a normal run.}
			\label{fig:main_experiment_normal}
		\end{subfigure}
		\hfill
		\begin{subfigure}{0.45\textwidth}
			\centering
			\includegraphics[width=\linewidth]{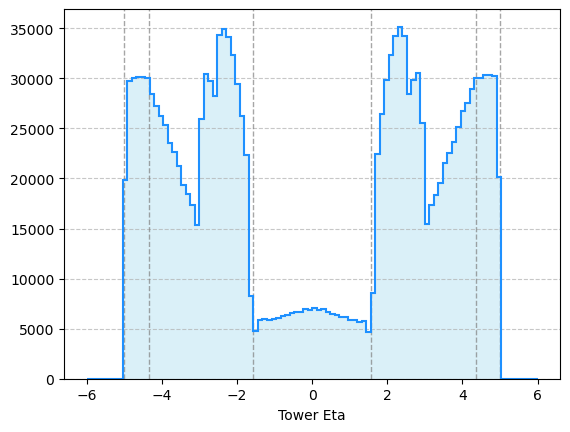}
			\caption{Distribution for HCAL barrel glitch.}
			\label{fig:main_experiment_barrel}
		\end{subfigure}
		
		\vspace{1em} 
		
		\begin{subfigure}{0.45\textwidth}
			\centering
			\includegraphics[width=\linewidth]{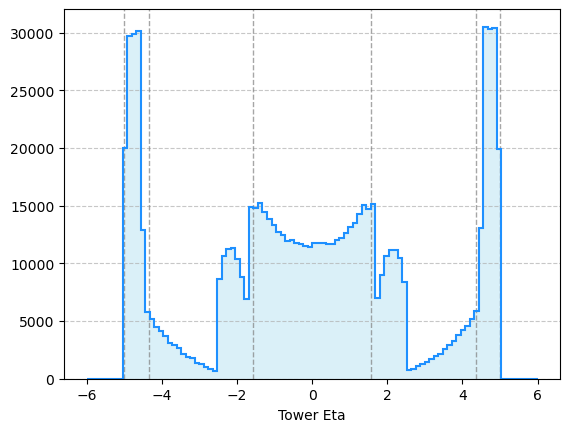}
			\caption{Distribution for HCAL endcap glitch.}
			\label{fig:main_experiment_endcap}
		\end{subfigure}
		\hfill
		\begin{subfigure}{0.45\textwidth}
			\centering
			\includegraphics[width=\linewidth]{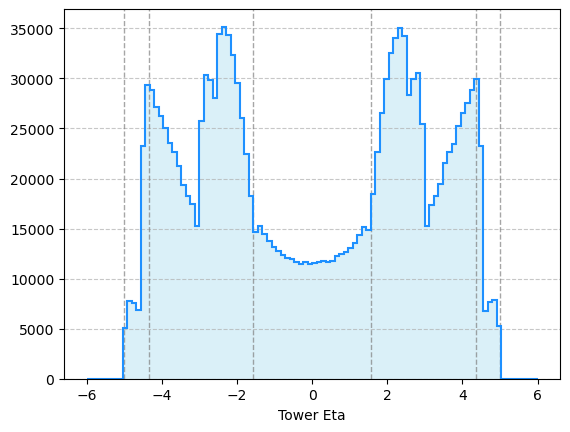}
			\caption{Distribution for HCAL forward glitch.}
			\label{fig:main_experiment_forward}
		\end{subfigure}
		
		\caption{Tower eta distribution for each of the $4$ detector simulation.}
		\label{fig:main_experiment}
	\end{figure}
	To generate training and validation data for the present work, we first simulated proton–proton collisions at a center-of-mass energy ($\sqrt{s}$) of $13$ TeV using \texttt{MadGraph5\_aMC@NLO} \cite{Alwall:2014hca}, with \texttt{Pythia8} \cite{Bierlich:2022pfr} for parton showering and hadronization. The output particle list was then passed to Delphes for detector-level simulation under the following four configurations:
	\begin{enumerate}
		\item \textit{Normal run:} all detector components active, which is also used as the reference run for further analysis.
		\item \textit{HCAL barrel glitch:} $5^\circ$ HCAL towers deactivated.
		\item \textit{HCAL endcap glitch:} $10^\circ$ HCAL towers deactivated.
		\item \textit{HCAL forward glitch:} $20^\circ$ HCAL towers deactivated.
	\end{enumerate}
	These controlled modifications, as shown in \cref{fig:main_experiment}, simulate distinct classes of detector malfunctions and together form the labeled dataset used to train and validate the MEDIC framework. Note that all four simulations, keep the ECAL unaltered and use the same set of particles generated through the \texttt{MadGraph5\_aMC@NLO} and \texttt{Pythia8} pipeline. This is there to ensure that the underlying \emph{physics process} is identical across scenarios, so MEDIC learns only the glitches in the detector and not differences in event generation process. Although our modification in Delphes \cite{arghya_delphes} allows each individual \emph{detector element} inside the ECAL and HCAL to be switched off, for this first proof of concept as well as for computational ease, we chose to deactivate sections of the detector as a single unit. The main inspiration of this approach is to demonstrate that the detector glitches can be realistically simulated and that such simulations can and should be incorporated into the development of ML based DQM tools.\\

	To emulate the continuous data-taking behavior of experiments, a \emph{sliding window dataset} is constructed over sequential detector events. A fixed window size $\mathcal{W}$ is chosen at the beginning of the dataset preparation, serving as a buffer that the network would use to process consecutive detector events. Each window therefore contains data from $\mathcal{W}$ successive events, including the observables mentioned in tables \ref{tab:tower_features}, \ref{tab:track_features} and \ref{tab:missinget_features} in appendix \ref{app:datas_create}. The window is advanced by one event at a time, adding the newest event and discarding the oldest one, thereby generating overlapping sequences that capture the temporal evolution of detector responses. This formulation effectively transforms the detector output into a time-series representation, reflecting how data accumulate in real-time operations. To ensure computational efficiency required for fast or online DQM, we only use partial information from each of the events by keeping $30$ randomly selected tracks (each with $7$ features) and $30$ randomly selected towers (each with $8$ features), during the construction of each window. This randomization introduces statistical diversity across windows while maintaining the lightweight nature required for real-time processing. The output label for each window corresponds to the empirical probability distribution of different detector states (normal, HCAL barrel glitch, HCAL endcap glitch, HCAL forward glitch) within that window. This design allows the learning framework to recognize both instantaneous and gradual anomalies while maintaining temporal continuity in the dataset. The details of the MEDIC network, trained on this dataset is explained in the following section while the technical details for labeling and preparing the training and testing dataset are mentioned in appendix \ref{app:datas_create}.

	\section{MEDIC: architecture and training}\label{sec:train_test}
	Following the discussions in the previous section and the algorithm outlined in Appendix \ref{app:datas_create}, one can construct a synthetic dataset to emulate continuous data taking in a collider experiment. Therefore, the next logical step is to explain the MEDIC network that can be used as a DQM tool using that dataset. But before discussing the details of the MEDIC network, it is important to address the heterogeneity of the detector outputs. Among the output branches generated by the Delphes simulations, three of them are used in this work, namely the tracks, calorimeter towers, and missing transverse energy (MET). Each of these branches comes with some features as explained in tables \ref{tab:tower_features}, \ref{tab:track_features} and \ref{tab:missinget_features} of appendix \ref{app:datas_create}. Tracks correspond to charged-particle trajectories reconstructed in the tracking system, calorimeter towers encode localized energy deposits in the ECAL and HCAL both combined together, and MET represents a global event-level quantity derived from the momentum imbalance. Because these quantities are not only conceptually distinct but also differ in structure and resolution, a network developed for DQM should process them through separate embedding pipelines before combining them in a unified representation for a classification task downstream. In the following, the structure of the MEDIC architecture is explained in detail.\\

	\begin{figure}[ht]
		\includegraphics[width=10cm]{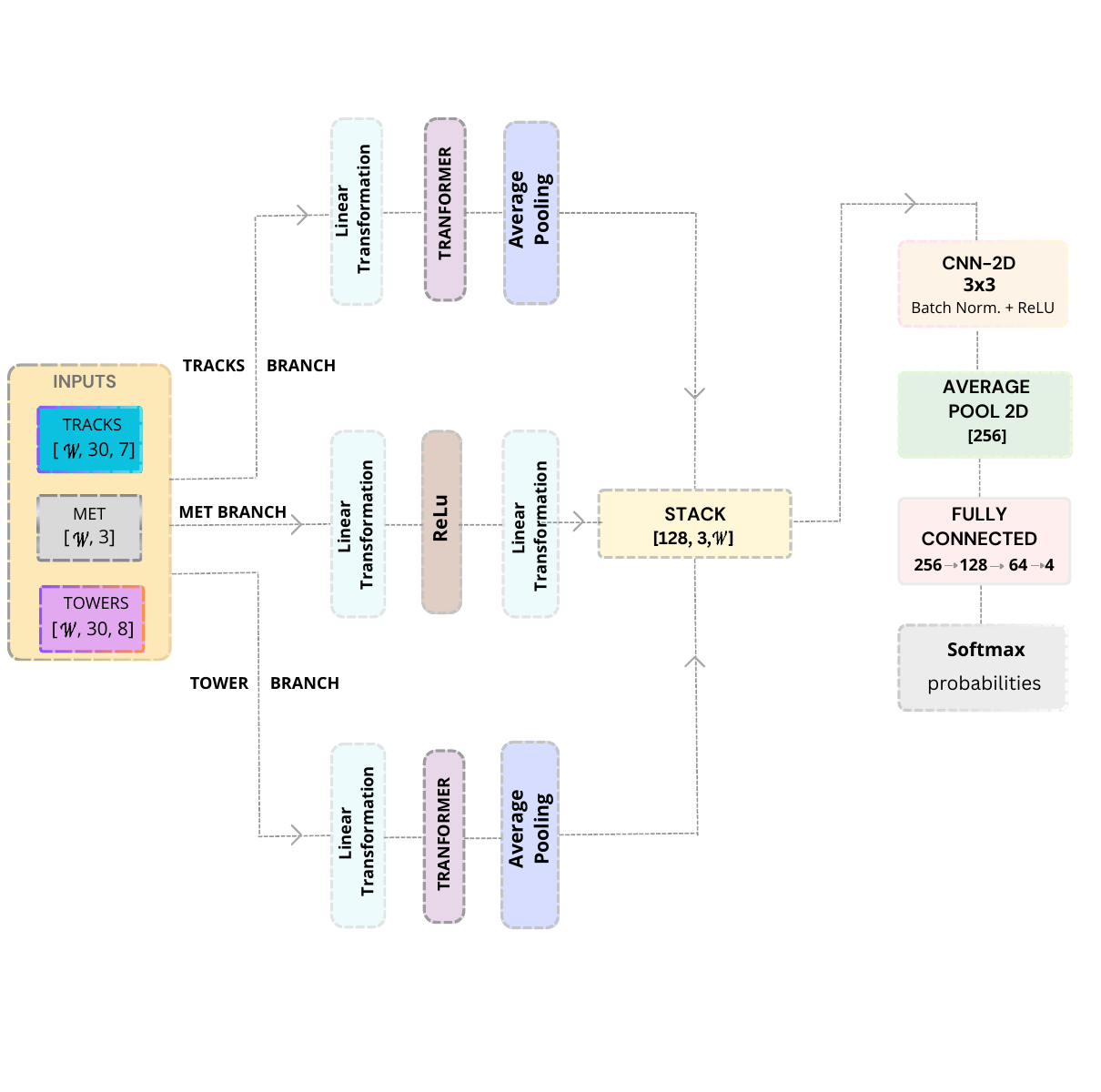}
		\centering
		\caption{Schematic representation of MEDIC neural network architecture. The model treats the three branches (Tracks, Towers, and MET) in three separate channels that encode detector inputs through linear projections, transformer encoders, and attention pooling. This is followed by a series of convolutional layers, then a global average pooling and a fully connected classifier to return probabilities. }
		\label{fig:MEDIC}
	\end{figure}

	The input structure of MEDIC is constrained by the choice of keeping only $30$ randomly selected tracks (with $7$ features) and towers (with $8$ features), a choice motivated by the aim of keeping the computational cost optimal and ensuring that MEDIC remains suitable for fast online monitoring. While the key parameters discussed above are carefully chosen for consistency, the remaining hyper-parameters for MEDIC are not extensively optimized and serve only as a reasonable baseline.

	\subsection{Architecture}
	Both the track and tower information sequences are inherently unordered sets of reconstructed objects, with no meaningful physical ordering even in the detector coordinates or in the $30$ track and tower information that are randomly selected. Therefore, to ensure permutation invariance, each collection is first projected into an embedding space via a linear layer and then processed by a transformer encoder based on multi-head self-attention \cite{vaswani2023attentionneed}. While self-attention is permutation-covariant (its output changes consistently under any reordering of the inputs), the permutation invariance is achieved by the subsequent attention-pooling layer, which assigns learnable weights to the embedded objects and aggregates them into a fixed length representation. This pooling step produces a vector of dimension $\left[\mathcal{W},128\right]$, where $128$ is simply a chosen hyperparameter for the embedding dimension. This design guarantees that the final representations are insensitive to any reshuffling of tracks or tower inputs, a crucial property for collider data where object ordering carries no physical meaning. Unlike track and tower features, MET details form a single event-level vector of dimension $3$ with energy, $\eta$ and $\phi$ information. Consequently, MET is processed through a  non-linear projection into the same embedding dimension of $128$ as the track and tower encoders.\\

	After obtaining the three embeddings, they are stacked into a tensor of shape $\left[128,3, \mathcal{W}\right]$, which can be interpreted as a three-channel feature map over the window $\mathcal{W}$ set of events. This structured representation is then passed through a deep $2D$ convolutional stack consisting of three convolutional blocks: a $3\times 3$ convolution with $64$ channels, followed by $3\times 3$ convolutions with $128$ and $256$ channels, respectively, each with batch-normalization and ReLU activation. These layers are meant to act as local feature extractors, allowing the network to learn correlations across the temporal dimension encoded in the collection of events inside the window $\mathcal{W}$. This stack of shape $\left[128, 3, \mathcal{W}\right]$, representing all the events and branches inside $\mathcal{W}$ is then followed by a global adaptive average pooling layer, that compresses the resulting activation map into a $256$-dimensional vector, which is fed into a fully connected classifier that produces the final \texttt{softmax} output with four nodes, one for each of the four detector quality probabilities.\\
	
	This architecture, shown in figure \ref{fig:MEDIC} therefore integrates permutation-invariant set encoders for tracks and towers, a dedicated embedding pathway for MET, and a convolutional module for classification. Therefore, from the standpoint of inference-time computational cost, MEDIC consists of a fixed per-event encoding stage followed by a window-level aggregation and classification stage. For each event within a window of size $\mathcal{W}$, the track and tower branches process a bounded number of objects ($30$ tracks with $7$ features each, $30$ towers with $8$ features each, and a $3$ dimensional MET vector) through linear projections, self-attention, and pooling, yielding a constant-cost event embedding of fixed dimension. These per-event embeddings, together with the MET branch, are stacked into a tensor of shape $\left[128,3, \mathcal{W}\right]$ (cf. the STACK layer in Fig.~\ref{fig:MEDIC}) and passed to the convolutional and pooling layers. As a result, the dominant inference-time cost scales linearly with the window size $\mathcal{W}$ while the per-event processing cost remains constant by construction. This scaling behavior provides a quick estimate of the order-of-magnitude computational resources  required when MEDIC is deployed for DQM inference. \\

	\subsection{Cost function \& Accuracy metrics}\label{subsec:accu}
	Before going into the details of training the MEDIC network, the input and output data structure of MEDIC is discussed here. As explained in the last section\footnote{confer appendix \ref{app:datas_create} for the algorithm.}, to simulate a continuous data-taking process, $\mathcal{W}$ number of event data are collected together as an input to the network, making the input as three heterogeneous branches, one for the tower with shape $\left[\mathcal{W}, 30, 8\right]$, another with track data with shape $\left[\mathcal{W}, 30, 7\right]$ and finally the last one with MET data with shape $\left[\mathcal{W}, 3\right]$. Since each event can come from any of the four performed simulations, one can associate each window with a probability vector with four entries. For example if one particular window has only $n_1$ normal entries, $n_2$ number of runs with HCAL barrel glitch, $n_3$ number of runs with HCAL endcap glitch and $n_4$ number of runs with HCAL forward detector glitch then the probability vector will just be\footnote{Note that $n_1+n_2+n_3+n_4=\mathcal{W}$ for the present case.} $\left[n_1/\mathcal{W},\,n_2/\mathcal{W},\, n_3/\mathcal{W},\, n_4/\mathcal{W}\right]$. This probability vector is the expected output from MEDIC, which can enable the shifter to assign a probability of anomaly as well as its source during the online data-taking process.\\

	Since the target of MEDIC is to learn a probability distribution, given the $\mathcal{W}$ sets of event data, a Kullback-Leibler (KL) divergence \cite{Joyce2011} between the predicted and target probability distribution is used as the cost function for training. Given the predicted log-probabilities  $\log p_{\theta}(y\,|\,x)$ and the target distribution 
	$q(y\,|\,x)$, where $x$ and $y$ denotes the input and output values and $\theta$ denotes the MEDIC hyper-parameters, the per–sample KL divergence used as cost function is defined as
	\begin{equation}\label{eq:loss}
		\mathrm{KL}(q \,\Vert\, p_{\theta})
		= \sum_{c=1}^{4} 
		q_c \log\!\left(
		\frac{q_c}{p_{\theta}(y|\,x)}
		\right),
	\end{equation}
	where the sum runs over all $4$ classes of probabilities.  In the present implementation of MEDIC \cite{medic}, the \texttt{PyTorch} \cite{10.1145/3620665.3640366} function $\texttt{F.kl\_div(log\_probs, target\_probs)}$ is used where it expects the first argument to be \emph{log-probabilities} and the second to be the \emph{target probabilities}.\\

	To better quantify the predictive performance of the network, in this work, two complimentary approaches are adopted. The first one is a \emph{hard} accuracy metric, based on the maxima of the predicted probability vector, whereas the second one is a \emph{soft modification} of Brier score, which compares the full probability distribution output by the classifier with the true or expected probability distribution. Accuracy reflects the discrete correctness of class assignments, by first assigning a \emph{true class} corresponding to each of the output $y_i$ as $c_i=\arg\max y_i$ so that the assigned class dictates the simulation class that dominates the events inside $\mathcal{W}$. Hence, one can define
	\begin{equation}\label{eq:acc}
		\mathrm{Acc} 
		= 
		\frac{1}{N} 
		\sum_{i=1}^{N}
		\mathbf{1}
		\!\left[
		\arg\max\, p_{\theta}(c\,|\,x_i)
		\;=\; c_i
		\right],
	\end{equation}
	where $c_i$ is the true class index and $c$ is the output of the network for a given data $x_i$ with $N$ being the total number of data points. While intuitive, this formulation of accuracy does not reflect the full capability of MEDIC network in assigning probabilities for each of the classes. To address this, a modified version of Brier score\footnote{Confer \cite{Brier1950} for the original formulation.} is also computed, which works as a scoring rule that measures the mean squared error between the predicted probability vector $p_{\theta}(c\,|\,x_i)$ and the target probability vector $y_i$ as 
	\begin{equation}\label{eq:brier}
		\text{Brier Score}= {1\over N}\sum_{i=1}^{N}\sum_{c=1}^{4} (p_{\theta}(c\,|\,x_i)- y_i)^2.
	\end{equation}
	Because the Brier score is sensitive to probability calibration, using both metrics in tandem increases robustness and provides a more reliable diagnostic for overfitting. By incorporating both a hard decision metric as accuracy in \eqref{eq:acc} and a soft, distribution-aware Brier score as defined in \eqref{eq:brier}, the model evaluation captures complementary information about the classifier’s performance. The accuracy metric ensures that the final decision boundaries are meaningful, while the Brier score tracks probabilistic reliability and model calibration.
	\begin{algorithm}[ht]
		\caption{$k$-Fold cross-validated training of MEDIC}
		\label{alg:medic_train}
		\begin{algorithmic}
			
			\State Stratify and construct $k$--fold partitions 
			$\{(\mathcal{D}^{(j)}_{\mathrm{train}},\,\mathcal{D}^{(j)}_{\mathrm{val}})\}_{j=1}^{k}$.
			
			\For{$j = 1$ to $k$} \Comment{Cross-validation loop}
			
			\State Initialize MEDIC model $M^{(j)}$.
			\State Compute class weights from $\mathcal{D}^{(j)}_{\mathrm{train}}$.
			\State Set $L_{\mathrm{best}}=\infty$ and patience counter $p=0$.
			
			\For{each epoch}
			\State \textbf{Training:} compute KL loss \eqref{eq:loss}, apply class weights, evaluate Brier Score \eqref{eq:brier} \State \hspace{1.6cm} and accuracy \eqref{eq:acc} over $\mathcal{D}^{(j)}_{\mathrm{train}}$, then update model parameters.
			
			\State \textbf{Validation:} Compute KL loss \eqref{eq:loss},  Brier score \eqref{eq:brier} and accuracy \eqref{eq:acc} over $\mathcal{D}^{(j)}_{\mathrm{val}}$. 
			\State $L_{\mathrm{val}}=$ KL loss over $\mathcal{D}^{(j)}_{\mathrm{val}}$.
			
			\If{$L_{\mathrm{val}} < L_{\mathrm{best}}$}
			\State Save parameters $M^{(j)}_{\mathrm{best}}$; reset $p=0$.
			\State $L_{\mathrm{best}} \gets L_{\mathrm{val}}$.
			\Else
			\State $p \gets p + 1$.
			\If{$p$ reaches patience threshold} \State \textbf{Early stop}. \State \textbf{break}
			\EndIf
			\EndIf
			
			\EndFor
			
			\EndFor
			
			\State Return all per–fold logs and the set of models $\{M^{(j)}_{\mathrm{best}}\}_{j=1}^k$.
			
		\end{algorithmic}
	\end{algorithm}

	\subsection{Cross validation and ensembles}
	As mentioned in the last section, some of the major constraints in the MEDIC network architecture come from the fact that we choose to keep $30$ randomly selected towers and tracks with respectively $7$ and $8$ features each. This decision is taken to make sure that the network input structure or the information-input to the model should not be unreasonably high, to keep the computational efficiency and cost in check. In the context of real-time  or online DQM for collider experiments, it is essential that the monitoring model remain robust against statistical fluctuations and variations in detector conditions. A single train/validation split is not sufficient for this application, as it may lead to over-estimation of performance and perhaps introduce unwanted sensitivity to the particular choice of the tracks and tower information as well as on the validation subset resulting in difficulties to assess the stability of the model. To overcome these limitations, in this case, a $k$-fold cross-validation strategy is adopted together with an ensemble evaluation over all trained folds. \\

	The basic idea of $k$-fold cross validation is to split the full data in $k$ number of equal parts or folds, then iteratively train a model $k$ times, each time keeping one of the $k$ parts of the data for validation and rest of the $k-1$ copies as the training data. This process is repeated $k$ times so that every sample appears exactly once in the validation set. To train the medic network we first kept $80\%$ of the full available data for training and validation and rest $20\%$ for testing. In the present case, an ensemble approach is taken in addition to a typical $k$- fold cross validation, where during the $k$ number of training cycles, different copies of the MEDIC network is trained independently. Hence, at the end of the training one have $k$ copies of MEDIC network, as part of a single ensemble. For inference during evaluation, the outputs of the ensemble (or $k$ copies of MEDIC network) are combined using two complementary strategies. For hard classification, a majority-vote rule is applied,
	\begin{equation}
		\hat{c}_{ens}(x)= \text{mode}\left\{\arg \max_{c}p_{\theta_j}(c|x)\right\}_{j=1}^k,
	\end{equation}
	where $\theta_j$ denotes the parameters of the $j$-th fold model.  The final accuracy is then computed as 
	\begin{equation}\label{eq:acc_ens}
		\mathrm{Acc}_{ens} 
		= 
		\frac{1}{N} 
		\sum_{i=1}^{N}
		\mathbf{1}
		\!\left[
		\hat{c}_{ens}(x_i)\;=\; c_i
		\right],
	\end{equation}
	where $c_i=\arg\max y_i$ with $(x_i,y_i)$ being the input and true-output pairs. On the other hand, for probabilistic evaluation, we first compute the ensemble-averaged predicted distribution,
	\begin{equation}
		\bar{p}(c|x)={1\over k}\sum_{j=1}^4 p_{\theta_j}(c|x),
	\end{equation}
	and then subsequently use the soft Brier score for the ensemble as 
	\begin{equation}\label{eq:brier_ens}
		\text{Brier Score}_{ens}= {1\over N}\sum_{i=1}^{N}\sum_{c=1}^{4} (\bar{p}(c\,|\,x_i)- y_i)^2.
	\end{equation}
	\subsection{Training}
	With the above components in place, the training procedure used for the MEDIC network can be summarized as follow. In this work, a $5$–fold cross-validation (or $k=5$) strategy is employed, resulting in an ensemble of five individually trained MEDIC models, each with its own set of parameters $\left\{\theta_j\right\}_{j=1}^5$ while the training is optimized stochastically using \texttt{Adam} \cite{Kingma2014AdamAM}. For every fold, batch size is fixed to be $64$ and a learning rate of $10^{-3}$ is chosen. To prevent overfitting, the training accuracy defined in \eqref{eq:acc} is continuously monitored, and early stopping is triggered once improvement saturates. The overall training workflow is provided in Algorithm~\ref{alg:medic_train}.

	\section{Results}\label{sec:results}
	With the architectural choices and training methodology described in \cref{sec:train_test}, several independent training runs were performed, each using a different window size $\mathcal{W}$. This scan allows one to identify the optimum choice of window length for which MEDIC achieves the best predictive performance under the previous choices for the hyper-parameters. Although MEDIC is trained on a \emph{four-class classification} task, the binary accuracies (between normal and anomalous runs) is also reported for completeness. By construction the model learns a full multi-class probability distribution, so the output probability mass is naturally spread across all four classes. Therefore, when evaluating binary performance, all the three anomalous classes are effectively collapsed into a single combined category. Since the binary thresholds are derived from these compressed probability masses, its binary performance is not optimized.  Therefore, for the binary case the performance is evaluated only through \emph{hard} predictions obtained via the accuracy metric \eqref{eq:acc} without the binary Brier score as the model does not produce a calibrated two class probability by construction.\\

	\begin{table}[ht]
		\centering
		\begin{tabular}{@{}c@{\hspace{1cm}}c@{\hspace{1cm}}c@{\hspace{1cm}}c@{\hspace{1cm}}c@{\hspace{1cm}}c@{}}
			\hline
			\multirow{2}{*}{\textbf{Window Size}} 
			& \multicolumn{3}{c}{\textbf{Multi-Class}} 
			& \multicolumn{2}{c}{\textbf{Binary}} \\
			& Acc & AUC & Brier 
			& Acc & AUC \\
			\hline
			$10$ &$0.836$ &$0.948$  &$0.007$  &$0.849$  &$0.926$   \\
			$20$ &$0.873$ &$0.957$  &$0.002$  &$0.883$  &$0.949$   \\
			$30$ &$0.897$  &$0.963$  &$0.001$  &$0.903$  &$0.961$   \\
			$40$ &$0.891$  &$0.962$  &$0.001$ &$0.901$  &$0.957$   \\
			$50$ &$0.898$  &$0.962$  &$0.001$ &$0.921$  &$0.957$   \\
			$60$ &$0.906$  &$0.964$  &$0.0008$ &$0.916$  &$0.959$   \\
			$70$ &$0.904$  &$0.964$  &$0.0006$ &$0.916$  &$0.956$   \\
			$80$ &$0.882$  &$0.962$  &$0.0007$ &$0.884$  &$0.954$   \\
			$90$ &$0.895$  &$0.962$  &$0.0006$ &$0.913$  &$0.954$   \\
			\hline
		\end{tabular}
		\caption{Performance metrics for different window sizes.}
		\label{tab:results}
	\end{table}

	Keeping the network architecture fixed, and employing the same early stopping method as outlined in Algorithm \ref{alg:medic_train}, the results for all tested values of $\mathcal{W}$ are summarized in \cref{tab:results}. A clear improvement is observed when increasing $\mathcal{W}$ from small values up to $\mathcal{W}=30$, indicating that very short windows do not provide sufficient contextual information for reliable classification. For larger window sizes, the performance metrics saturate as accuracy, AUC, and calibration remain stable for $\mathcal{W}>30$ with no systematic improvement or degradation observed up to $\mathcal{W}=90$.  We therefore choose $\mathcal{W}=30$ for subsequent analyses as the smallest window size at which near-maximal performance is achieved. This choice balances classification performance with practical considerations, since larger windows increase training and inference cost without providing consistent performance gains. Larger window sizes primarily demonstrate the robustness of the approach rather than defining a distinct optimal regime. Consequently, for all subsequent discussions including full training curves and confusion matrices, we focus exclusively on the choice $\mathcal{W} = 30$, to illustrate the performance and robustness of the MEDIC framework.\\

\begin{figure}[ht]
		\centering
		\begin{subfigure}[t]{0.49\textwidth}
			\centering
			\includegraphics[width=\linewidth]{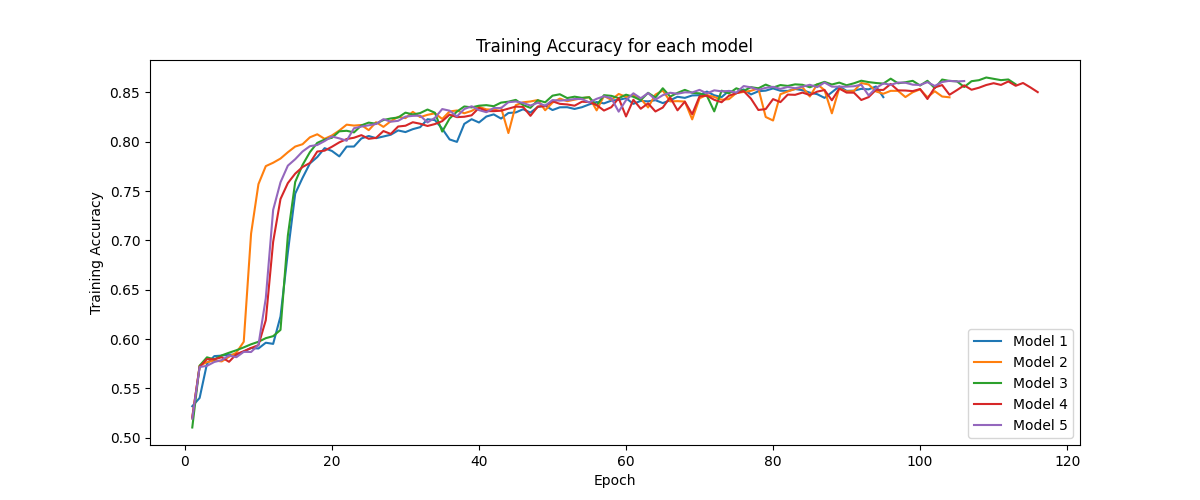}
			\caption{Training Accuracy}
		\end{subfigure}
		\hfill
		\begin{subfigure}[t]{0.49\textwidth}
			\centering
			\includegraphics[width=\linewidth]{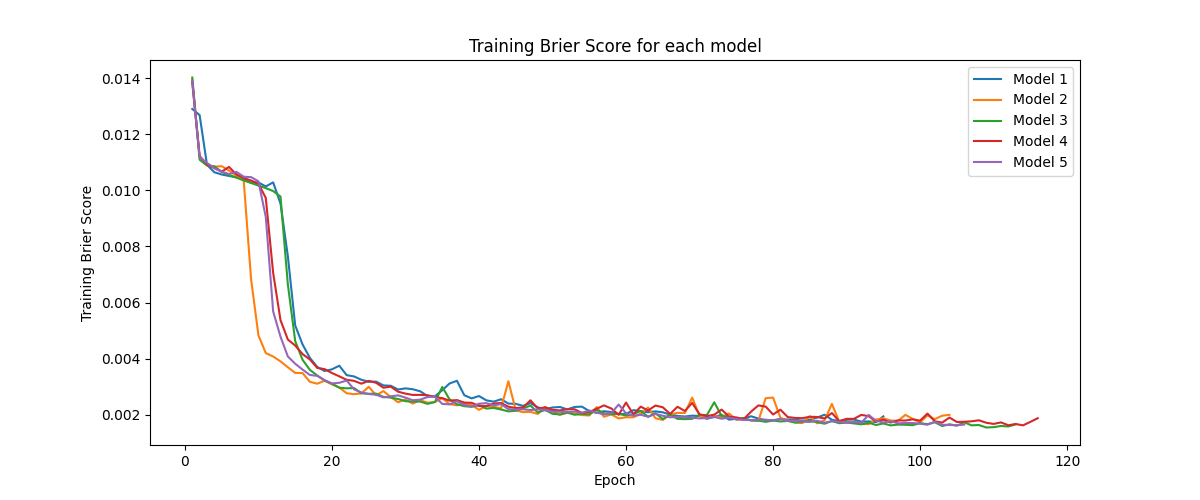}
			\caption{Brier Score}
		\end{subfigure}
		
		\vspace{0.5cm}
		
		\begin{subfigure}[t]{0.5\textwidth}
			\centering
			\includegraphics[width=\linewidth]{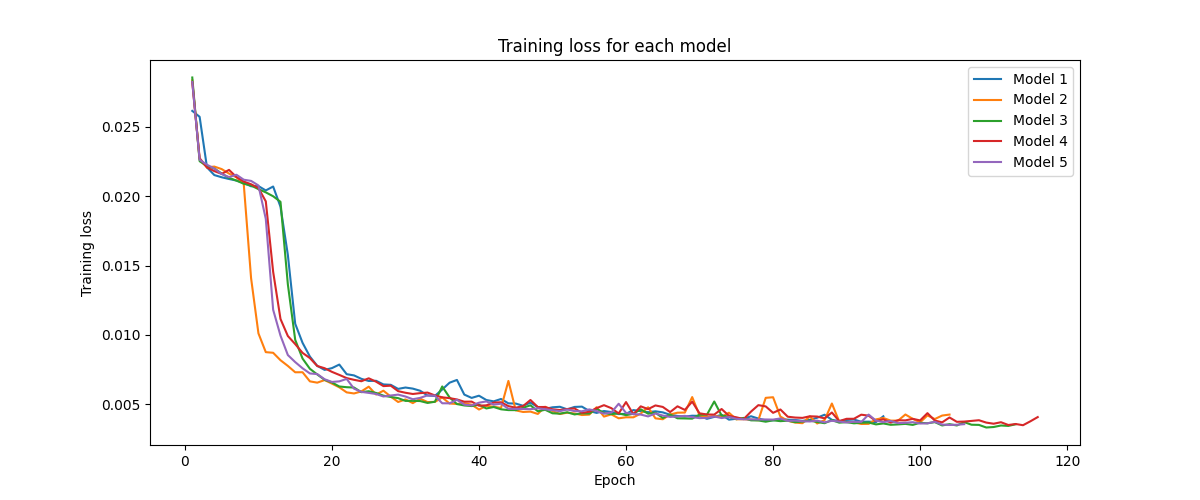}
			\caption{KL divergence Loss}
		\end{subfigure}
		
		\caption{Training metrics for $\mathcal{W}=30$ across all epochs and $5$ folds.}
		\label{fig:training_metrics}
	\end{figure}

	All the experiments were performed on a Linux server featuring dual AMD EPYC $7313$ processors ($64$ threads) and eight NVIDIA RTX A5000 GPUs ($24$ GB GDDR6 memory). All training used CUDA-accelerated \texttt{PyTorch}. For $\mathcal{W}=30$, it took approximately $16$ hours to complete the training with early stopping enabled. The evaluation of metrices during the training process is shown in figure \ref{fig:training_metrics}. While the metrics reported in \cref{tab:results} quantify training and validation performance, MEDIC is intended to be used purely in inference mode during actual DQM operation. Therefore the relevant computational cost is determined by the per-window inference time and memory footprint. Since events inside a window are processed independently by the track, tower, and MET encoders before aggregation (cf. the STACK layer in \cref{fig:MEDIC}), the inference time scales approximately linearly with the window size $\mathcal{W}$, while the memory usage scales linearly with both $\mathcal{W}$ and the fixed number of selected tracks and towers per event. With the input dimensionality capped at a finite number of tracks and towers, which is chosen to be $30$ for the present analysis, the per-event processing cost remains bounded, making the approach compatible with fast or online DQM workflows. This inference-time scaling, rather than the offline training cost, is the relevant consideration for practical deployment.\\

\begin{figure}[ht]
		\centering
		\begin{subfigure}[t]{0.48\textwidth}
			\centering
			\includegraphics[width=\linewidth]{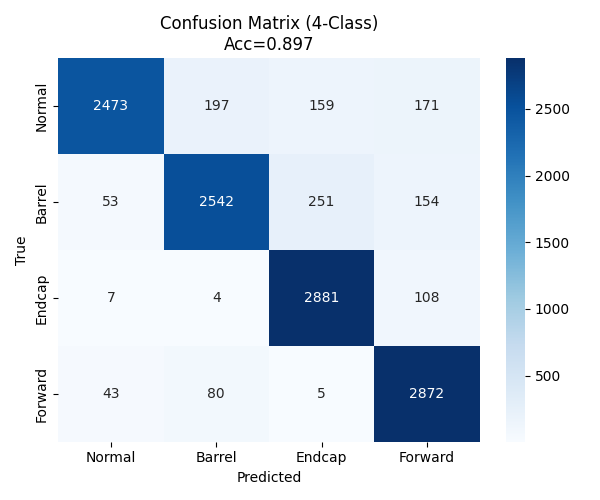}
			\caption{Multi-Class Confusion Matrix}
		\end{subfigure}
		\hfill
		\begin{subfigure}[t]{0.48\textwidth}
			\centering
			\includegraphics[width=\linewidth]{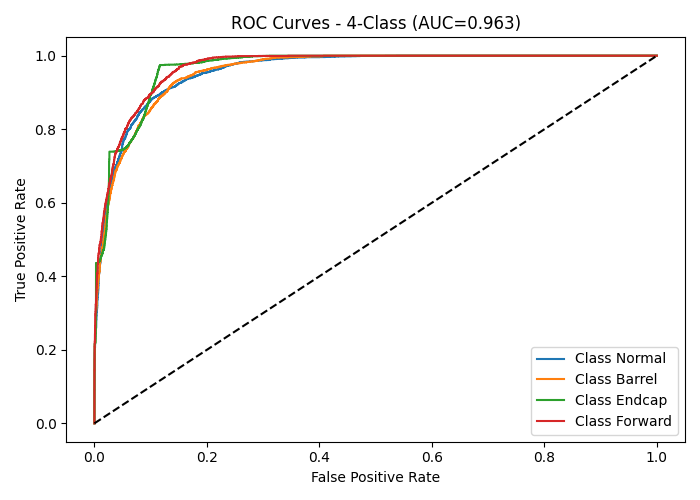}
			\caption{Multi-Class ROC-AUC (One-vs-Rest)}
		\end{subfigure}
		\vspace{0.5cm}
		
		\begin{subfigure}[t]{0.48\textwidth}
			\centering
			\includegraphics[width=\linewidth]{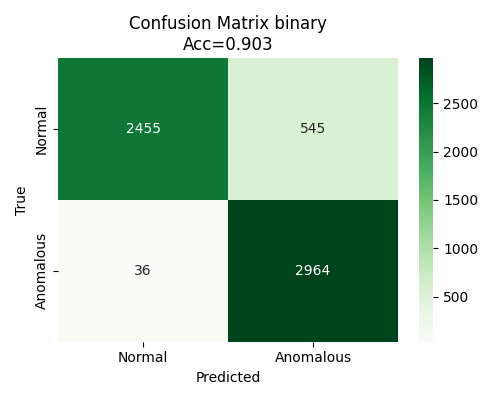}
			\caption{Normal vs Anomalous Confusion Matrix}
			\label{fig:confusion_binary}
		\end{subfigure}
		\hfill
		\begin{subfigure}[t]{0.48\textwidth}
			\centering
			\includegraphics[width=\linewidth]{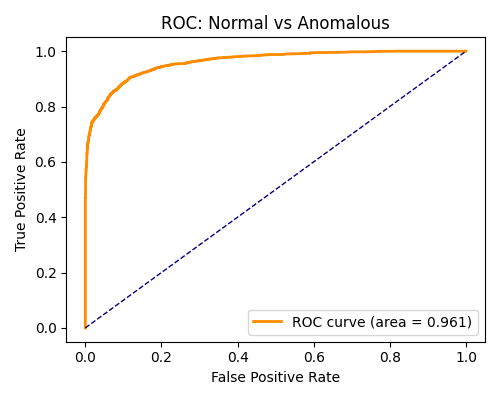}
			\caption{Normal vs Anomalous ROC-AUC}
		\end{subfigure}
		
		\caption{Evaluation of MEDIC on test data for $\mathcal{W}=30$ with top row being multi-class performance and bottom row demonstrating binary performance.}
		\label{fig:multi_class_binary}
	\end{figure}

	The confusion matrices and ROC-AUC curves are summarized in figure \ref{fig:multi_class_binary}. For the multi-class case, hard predictions as defined in \eqref{eq:acc} is implemented for the confusion matrix and a one-vs-rest approach for ROC-AUC, taking $3000$ samples per class from the test set to get a balanced dataset. For the binary case (normal vs. anomalous), $3000$ normal and $3000$ anomalous samples are used. Although MEDIC is not explicitly tuned for binary classification, it still achieves strong performance with most normal and anomalous events correctly identified. The small number of misclassifications reflects the collapsed probability mass from the three anomalous classes into one as mentioned before, yet the network still effectively separates normal from anomalous data.
	In light of the above results one can observe that MEDIC effectively captures the \emph{temporal} patterns (as mimicked by the sliding window $\mathcal{W}$) in the input features and provides robust multi-class predictions, with $\mathcal{W}=30$ yielding the best overall performance. Even when evaluated in a simplified binary setting, it maintains strong discriminative power. These findings confirm that the architectural choices and training procedure, including early stopping and class balancing as described in the training algorithm \ref{alg:medic_train}, are well-suited for both multi-class and binary classification tasks. Together, they demonstrate that MEDIC is well suited as a viable and practical DQM method.\\

	To relate the reported performance metrics and accuracy information to practical DQM usage, the network output can be interpreted as an anomaly score evaluated for each sliding window. In an operational setting, alerts would not be triggered by a single anomalous window, but only when $\mathcal{N}$ consecutive windows exceed a chosen anomaly threshold. This requirement helps avoid reacting to short-lived fluctuations while still allowing persistent detector issues to be identified. For example, the binary confusion matrix in \cref{fig:confusion_binary} indicates a per-window false-positive rate of $\approx 18\%$, which will be reduced exponentially when several consecutive anomalous windows are taken into consideration. In practice, the value of $\mathcal{N}$ would be chosen by the experiment to balance sensitivity against the acceptable false-alert rate, with larger $\mathcal{N}$ reducing accidental alerts at the cost of a longer response time, consistent with a human-in-the-loop DQM procedures. The high AUC and favorable calibration observed for $\mathcal{W}=30$ ensure that such thresholds can be adjusted according to experiment-specific requirements.

	\section{Discussion and Conclusion}\label{sec:discussion}
	In this paper, a novel simulation-driven approach to DQM in collider experiments is presented along with the  introduction of an end-to-end network, MEDIC, capable of identifying and classifying different types of detector glitches. Our choice of hyper-parameters and input size was guided by the requirement for a fast or online DQM, where minimizing processing time is crucial while still achieving high accuracy. By using simulation-based data, even before real experimental data becomes available, this approach enables early development, testing, and optimization of DQM pipelines. Importantly, the reference run used for training does not need to be manually certified by experts, which significantly reduces the human overhead in setting up the monitoring systems or in this case the ML models. While the paper focuses on collider example, this methodology can be applied on other particle detector experiments as well after appropriate adaptations.\\

	The advantage of this simulation-driven approach lies in its flexibility. One can simulate a variety of detector conditions, which allows ML models to learn robust representations of anomalous behaviors. In the present study, MEDIC is trained on particle-level kinematics obtained from Delphes simulations \cite{deFavereau:2013fsa}. Even though only large-scale detector components were switched off in the scenarios presented here, the data was generated such that any temporal window could contain any combination of the four simulated situations. Notably, MEDIC was not explicitly trained for binary classification between normal and anomalous runs, nonetheless, it demonstrates strong predictive performance for that as well.\\

	It should be emphasized that Delphes provides only particle-level outputs, not detailed electronic-level signals. Its digitization process is inherently limited given its modular and fast-simulation framework that relies on parameterization of the overall detector response instead of simulating the underlying physics processes in detail. This simplification limits the resolution and number of features available to the network, which occasionally leads to ambiguities between normal and anomalous events as can be observed from the confusion matrices in figure \ref{fig:multi_class_binary}. A more detailed simulation using Geant4 \cite{GEANT4:2002zbu}, would likely improve the model accuracy as Geant4 performs a detailed simulation with precise position and timing information of energy deposits. It also converts raw energy deposits into realistic detector signals, accounting for specific hardware effects coming from signal generation, transport and electronic response combined with detector resolution effects. This translates into more information rich features using Geant4 that might reflect in improved accuracy. Despite this limitations, our results indicate that a ML model trained solely on Delphes-level data can still successfully detect anomalies and even identify their sources.\\

	This work demonstrates the potential of simulation-driven DQM as a new paradigm for collider experiments. Any experiment can benefit from generating simulated reference runs and glitch scenarios prior to data-taking runs, enabling the development and validation of automated monitoring tools in advance. Although the current result is promising and a more detailed simulation with electronic input and perhaps with more data points might increase the accuracies of prediction, our approach is still intended as a human-in-the-loop system, where shifters are assisted by the network’s predictions rather than fully replaced. In this framework, alerts are issued only after persistent anomalous behavior is observed over multiple consecutive windows, ensuring that human attention is drawn to sustained issues rather than simple statistical fluctuations. Such a pipeline allows for rapid anomaly detection and classification, providing actionable insights in real time and significantly aiding detector operations. This would be particularly valuable in the context of upcoming HL-LHC runs with their extremely high data volumes, where automated tools such as MEDIC could significantly enhance the efficiency of the shifter by rapidly flagging anomalies along with pointing out the source of error, thereby significantly reducing the manual load of real-time DQM.\\

	While the present study focuses on a limited number of labeled detector configurations as a proof of concept, the MEDIC framework is intended to be extended to a wider range of detector conditions, including partially labeled or unlabeled scenarios. In a realistic DQM deployment, this does not require a single classifier with a very large number of output classes. Instead, the framework can scale naturally through hierarchical or modular strategies, where deviations from nominal detector behavior are detected first and then classified at the level of subsystems or detector regions. A key feature of this approach is that the training data are generated entirely through simulation by explicitly introducing controlled detector fault scenarios, rather than relying only on statistical fluctuations in real data, which distinguishes MEDIC from many traditional DQM methods.\\

	In this spirit, MEDIC can be viewed as a modular architecture in which shared event-encoding layers are reused, while only the final classification head is adapted or replaced as new detector scenarios are introduced. This allows multiple binary or few-class classifiers, guided by historical knowledge of common failure modes, to operate in parallel and be extended incrementally without retraining the full network. The same framework can also be adapted to unsupervised or semi-supervised settings by learning a representation of nominal detector behavior and flagging persistent deviations in the latent space, enabling sensitivity to unforeseen failure modes. As with any simulation-driven approach, network performance may be affected by inaccuracies in detector modeling; a systematic study of such effects using dedicated Monte Carlo variations is therefore an important direction for future work and beyond the scope of the present study. The implementation associated with this work is publicly available in the MEDIC GitHub repository \cite{medic} and is provided to support transparency and reproducibility of the results presented here. The repository currently reflects an active research codebase, and future methodological developments discussed in this work will be incorporated within the same framework.\\

	Overall, the results underscore the robustness and practical utility of MEDIC for fast DQM. The methodology presented here lays the groundwork for future extensions, including full electronic-level simulations and integration with real detector outputs, while highlighting that even simplified particle-level simulations can produce valuable tools and insights for operational monitoring in high-energy physics experiments.

	
	\section*{Acknowledgment}
	\noindent This work is supported by U.S.A. National Science Foundation Award OAC-2334265. We would like to thank Roy Cruz (UPR Mayag\"uez), Tetiana Mazurets (UPR Mayag\"uez) and Gabriele Benelli (Brown University) for illuminating discussions. 
	
	
	\appendix
	\section{Dataset Generation}\label{app:datas_create}
	The first step in the data generation pipeline for MEDIC consisted of using \texttt{MadGraph5\_aMC@NLO} \cite{Alwall:2014hca} to generate $100,000$ proton-proton collision events to produce two jets with $\sqrt{s}=13$ TeV. The resulting file was then passed through \texttt{Pythia8} \cite{Bierlich:2022pfr} for parton showering and hadronization. Then Delphes \cite{deFavereau:2013fsa, arghya_delphes} is used to simulate the 4 distinct CMS detector responses as mentioned at the end of \cref{sec:det_sim} (corresponding $\eta$ distributions are shown in \cref{fig:main_experiment}). \\
	\begin{algorithm}[ht]
		\caption{\textbf{
				Dataset Creation with Randomized Event Sampling for a fixed window size}}
		\label{alg:sliding_window_dataset}
		\begin{algorithmic}
			\Require Event list $\mathcal{E} = \{E_1, E_2, \dots, E_{N_e}\}$, window size $\mathcal{W}$
			\Require number of tracks $N_{\text{track}}=30$, number of towers $N_{\text{tower}}=30$
			\State Randomly permute the events: $\mathcal{E} \gets \text{Shuffle}(\mathcal{E})$
			\State Compute number of windows: $N_\mathcal{W}= N_e - \mathcal{W} + 1$
			\State Initialize empty lists: $X_{\text{tracks}}$, $X_{\text{towers}}$, $X_{\text{miss}}$, $y$
			\vspace{3pt}
			\For{$i = 1$ to $N_\mathcal{W}$} \Comment{Slide the window across events}
			\State Initialize empty lists: $\mathcal{T}_{\text{set}}, \mathcal{H}_{\text{set}}, \mathcal{M}_{\text{set}}, \mathcal{P}_{\text{set}}$
			\For{$j = i$ to $i + \mathcal{W} - 1$} \Comment{Iterate through events in window}
			\State $E_j \gets \mathcal{E}[j]$ 
			\vspace{4pt}
			\State \textbf{Randomize track selection:}
			\State $n_t \gets |\text{tracks}(E_j)|$
			\If{$n_t > 0$}
			\State Select random indices $I_t \sim \text{Uniform}([1, n_t], \min(n_t, N_{\text{track}}))$
			\State $T_j \gets \text{tracks}(E_j)[I_t]$
			\Else
			\State $T_j \gets \text{ZeroMatrix}(N_{\text{track}}, d_t)$
			\EndIf
			\State Pad $T_j$ to shape $(N_{\text{track}}, d_t)$ and append to $\mathcal{T}_{\text{set}}$
			
			\vspace{4pt}
			\State \textbf{Randomize tower selection:}
			\State $n_h \gets |\text{towers}(E_j)|$
			\If{$n_h > 0$}
			\State Select random indices $I_h \sim \text{Uniform}([1, n_h], \min(n_h, N_{\text{tower}}))$
			\State $H_j \gets \text{towers}(E_j)[I_h]$
			\Else
			\State $H_j \gets \text{ZeroMatrix}(N_{\text{tower}}, d_h)$
			\EndIf
			\State Pad $H_j$ to shape $(N_{\text{tower}}, d_h)$ and append to $\mathcal{H}_{\text{set}}$
			
			\vspace{4pt}
			\State \textbf{Missing energy:} extract $M_j = \text{missingET}(E_j)$ and append to $\mathcal{M}_{\text{set}}$
			\State \textbf{Blindfold label:} append $b_j = \text{blindfold}(E_j)$ 
			\State Compute window-level output probabilities:
			\[
			\begin{matrix}
				p_1^{\text{(normal)}}=\frac{1}{\mathcal{W}}\sum_{j=1}^{\mathcal{W}}b_j^{(\text{normal})}; & p_2^{\text{(barrel glitch)}}=\frac{1}{\mathcal{W}}\sum_{j=1}^{\mathcal{W}}b_j^{(\text{barrel glitch})} \\
				p_3^{\text{(endcap glitch)}}=\frac{1}{\mathcal{W}}\sum_{j=1}^{\mathcal{W}}b_j^{(\text{endcap glitch})}; & p_4^{\text{(forward glitch)}}=\frac{1}{\mathcal{W}}\sum_{j=1}^{\mathcal{W}}b_j^{(\text{forward glitch})}
			\end{matrix}
			\]
			\State \textbf{Probabilities:} append $p_i$ with $i\in[1,4]$ to $\mathcal{P}_{\text{set}}$
			\EndFor
			
			\vspace{4pt}
			
			\State Append $\mathcal{T}_{\text{set}}, \mathcal{H}_{\text{set}}, \mathcal{M}_{\text{set}}, \mathcal{P}_{\text{set}}$ to datasets $X_{\text{tracks}}$, $X_{\text{towers}}$, $X_{\text{miss}}$, $y$
			\EndFor
			
			\vspace{4pt}
			\State \Return $(X_{\text{tracks}}, X_{\text{towers}}, X_{\text{miss}}, y)$
			
		\end{algorithmic}
	\end{algorithm}

	\begin{table}[ht]
		\centering
		\begin{tabular}{|c|c|}
			\hline
			Feature         & Description \\
			\hline 
			Tower.ET        & calorimeter tower transverse energy \\
			Tower.Eta       & calorimeter tower pseudorapidity \\
			Tower.Phi       & calorimeter tower azimuthal angle \\
			Tower.E         & calorimeter tower energy \\
			Tower.T         & calorimeter deposit time, averaged over all particles \\
			Tower.NTimeHits & number of hits contributing to time measurement \\
			Tower.Eem       & calorimeter tower electromagnetic energy \\
			Tower.Ehad      & calorimeter tower hadronic energy \\
			\hline
		\end{tabular}
		\caption{Delphes tower features.}
		\label{tab:tower_features}
	\end{table}
	\begin{table}[ht]
		\centering
		\begin{tabular}{|c|c|}
			\hline
			Feature        & Description   \\
			\hline 
			Track.P        & track momentum \\
			Track.PT       & track transverse momentum \\
			Track.Eta      & track pseudorapidity \\
			Track.Phi      & track azimuthal angle \\ 
			Track.CtgTheta & track cotangent of theta \\
			Track.C        & track curvature inverse \\
			Track.L        & track path length \\
			\hline
		\end{tabular}
		\caption{Delphes track features.}
		\label{tab:track_features}
	\end{table}
	\begin{table}[ht]
		\centering
		\begin{tabular}{|c|c|}
			\hline
			Feature       & Description \\
			\hline 
			MissingET.MET & missing transverse energy \\
			MissingET.Eta & missing energy pseudorapidity \\
			MissingET.Phi & missing energy azimuthal angle \\
			\hline
		\end{tabular}
		\caption{Delphes missinget features.}
		\label{tab:missinget_features}
	\end{table}

	Of the four resulting ROOT files, each containing seventeen branches, three were kept with select features as mentioned in tables \ref{tab:tower_features}, \ref{tab:track_features} and \ref{tab:missinget_features}. A fourth branch was then manually added with the purpose of containing four labels, each corresponding to a detector response: normal run, HCAL barrel glitch, HCAL endcap glitch, HCAL forward glitch. For example, an event produced from the simulation in which the HCAL barrel was turned off (in other words an event with barrel glitch) would be identified with $\left[0, 1, 0, 0\right]$. This way of labeling keeps the possibility open for generating events with multiple simultaneous detector errors in future studies. \\

	Having extracted the desired features and equipped it with the label branches, the data was then concatenated along the event axis and shuffled. This results in a data file containing $80,000$ events ($20,000$ events for each simulation). One should note that, the underlying physics events generated using the same \texttt{MadGraph5\_aMC@NLO} \texttt{+} \texttt{Pythia8} chain remain identical across all four simulations, ensuring that any differences arise solely from detector level effects. A filtering was then carried out for convenience in which all events containing less than $10$ entries in the track or tower branches were removed from the data. Finally, to emulate the continuous data-taking behavior of collider experiments as explained in \ref{sec:det_sim}, we construct a sliding window dataset to collect $\mathcal{W}$ number of events together. The probability vector associated with each window is defined as discussed in \cref{subsec:accu}. The pseudo-code for simulating this \emph{time-series} data is shown in Algorithm \ref{alg:sliding_window_dataset}, which receives the refined event-level data and returns the time series or window-level data with the probabilities mentioned in \cref{sec:det_sim}. 
	
	
	\bibliographystyle{JHEP}  
	\bibliography{biblio} 
\end{document}